# Influence of vacancies on phase transition in an organic molecular crystal

M. A. Korshunov[1]

*Kirensky Institute of Physics, Siberian Division, Russian Academy of Sciences, Krasnoyarsk, 660036 Russia*

**Abstract.** By the example of phase transition (α - β) in p-dichlorobenzene it is shown, that for phase transition presence of vacancies is necessary. In an ideal crystal, such transition is improbable, that is confirmed with calculations and their consent with experimental information on spectra of the lattice oscillations.

In the molecular crystals, phase transitions flow past as against nuclear structures not only at displacements of centre of gravity, but also change of orientation of molecules. These changes will happen, if it is energy possible. As is known actual crystals contain flaws (vacancies, impurities, etc.). There is a problem, whether phase transition if the structure of a crystal is ideal is possible.

For clearing up of this problem calculation of energy of a lattice of the molecular crystal, both with ideal structure have been carried out, and at presence in structure of vacancies. Thus, displacements of centre of gravity and change of orientation of molecules happening was taken into account at phase transition.

As object of examination, p-dichlorobenzene has been chosen. Various authors repeatedly explored P-dichlorobenzene. It has been found, that at temperature 30.8 °C there is a polymorphic transmutation. The low-temperature monoclinic α - a phase with two molecules in a unit cell transfers in high-temperature триклинную β - a phase with one molecule in a unit cell [1].

Calculations of interaction of molecules were carried out, using a method atom - atom of potentials [2]. Calculations were carried out for lines of temperatures.

At phase, transition of a molecule should be unfolded, but as has shown calculation, in case of an ideal crystal such rotational displacement is energy impossible even if to use parameters of a lattice the close to temperature of a fusion. Energy of a lattice in β - a phase of p-dichlorobenzene with ideal structure makes 14.58 kcal / mole in α - a phase 15.53 kcal / mole.

In case of an ideal lattice at transition from β - phases in α - a phase gyration of a molecule around of axes with the greatest and medial moment of inertia are energy possible for rotational displacements on corners no more than $20^0$. For an axis with the least moment of inertia no more than $40^0$. That is necessary corners for rotational displacement of molecules at phase transition will not be reached. It, apparently, shows that in an ideal molecular crystal of p-dichlorobenzene

---

[1] E-mail: mkor@iph.krasn.ru

polymorphic transition is improbable.

Therefore, calculation of rotational displacements of molecules in a crystal of p-dichlorobenzene containing vacancies has been carried out. Appeared that for molecules located about vacancy rotational displacement of molecules as is possible at transition from α in β a **phase** and on the contrary. There is a rotational displacement and displacement of a viewed molecule that is inconvenient to ideal structure. Thus, parameters of a lattice are close to what correspond to structure at 33ºC.

For acknowledgement, that in a real crystals there are defects (vacancies) which, apparently, promote phase transition, spectrums of small frequencies of p-dichlorobenzene in β - a phase (figure) have been obtained.

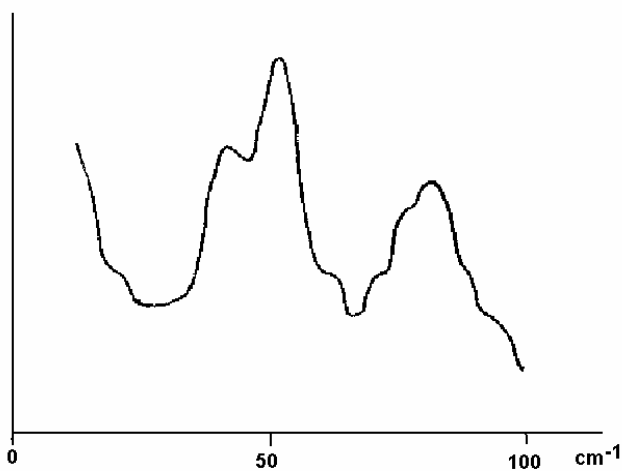

Fig. A spectrum of the lattice oscillations β - phases of p-dichlorobenzene.

In an ideal crystal it should be observed three intensive lines, the bound with orientation oscillations of molecules. In a spectrum of small frequencies β - p-dichlorobenzene a series of additional lines of small intensity has been found. In a spectrum α - phases as a series of additional lines of small intensity, which occurrence as shown in operation [3] is observed is stipulated by presence of vacancies in structure. For an explanation of additional lines in a spectrum β - p-dichlorobenzene calculations of frequencies of the lattice oscillations are carried out. Calculations of spectrums were carried out on a method the Dyne [4] . This method allows to carry out calculation of frequency spectrums, for ideal structures, and random structures.

Calculation of a spectrum of an ideal molecular crystal of p-dichlorobenzene in β - a phase has been at first carried out. It is obtained three values of frequencies of lines relevant to experimental data. At phase transition there is a gyration of molecules on above given values of corners. The spectrums of the lattice oscillations found at these parameters do not correspond to the experimental spectrums.

Then calculation of a histogram of frequencies has been lead at presence in structure of vacancies. The frequencies of the lattice oscillations designed at it are close to the experimental effects as in α and β phases. Calculations have shown, that occurrence of additional lines is stipulated by presence of vacancies which presence calls orientation disorder of molecules environmental vacancy.

Hence, it set presence of

vacancies in an actual crystal of p-dichlorobenzene having polymorphic transition.

Thus, as show calculations for polymorphic transition from α in β a phase transition in structure of p-dichlorobenzene of vacancies is necessary. In case of ideal structure such transition is improbable.